\documentclass[twocolumn,prl]{revtex4}

\usepackage{amssymb}
\usepackage{graphicx}

\begin{document}

\title{Growth of single wall carbon nanotubes from $^{13}$C isotope labelled
organic solvents inside single wall carbon nanotube hosts}
\author{Ferenc Simon $^{\ddag }$}
\author{Hans Kuzmany}
\affiliation{Institut f\"{u}r Materialphysik, Universit\"{a}t Wien,
Strudlhofgasse 4, A-1090 Wien, Austria}

\begin{abstract}
Exploring the synthesis of novel molecular nanostructures has been
in the forefront of material research in the last decade. One of the
most interesting nanostructures are single wall carbon nanotubes
(SWCNTs). Their catalyst free growth, however, remains an elusive
goal. Here, we present the growth of single wall carbon nanotubes
from organic solvents such as benzene and toluene in a confined
environment, inside a host SWCNT. The solvents encapsulated in
SWCNTs are transformed to an inner tube when subject to a heat
treatment under dynamic vacuum at 1270 $^{\circ}$C. We used isotope
labeling of the different carbon sources to prove that the source of
the inner tubes is indeed the solvent. Our results put constraints
on the models explaining the inner tube growth and provides a simple
alternative for the fullerene based inner tube growth. It also
provides the possibility to study a completely new field of
in-the-tube chemistry.
\end{abstract}



\maketitle

\section{Introduction}

Catalyst free growth of single wall carbon nanotubes (SWCNT) has
been intensively attempted since their discovery in 1993 \cite%
{IijimaNAT1993,BethuneNAT1993}. The efforts resulted in novel
nanostructures such as e.g. the nano-horns \cite{IijimaCPL1999} but
the metal catalyst-free synthesis remains elusive. Recently,
catalyst free growth of SWCNTs was achieved from fullerenes
encapsulated in SWCNTs, when these so-called peapods \cite{SmithNAT}
are subject to a high temperature annealing at 1270 $^{\circ}$C
\cite{BandowCPL2001,PfeifferPRL2003}.

In this paper the synthesis of inner tubes is reported from $^{13}$C
isotope labeled organic solvents such as benzene and toluene
encapsulated in SWCNTs. This is demonstrated to work when fullerenes
are co-encapsulated preventing the solvents from evaporation.
Isotope labeling proves unambiguously that the solvents contribute
to the inner tubes and gives a measure on the yield. The current
result opens new perspectives for the catalyst free synthesis of
SWCNTs in other confined environments such as zeolite
\cite{TangSCSCI} and allows further exploration of the in-the-tube
chemistry.

\section{Experimental}
\textit{Sample preparation.} Commercial SWCNT material (50 weight \%
purity, Nanocarblab), fullerenes of natural carbon (Hoechst AG),
benzene and toluene (Sigma Aldrich) and isotope labeled benzene and
toluene (Euriso-Top SA) were used. The purification of the SWCNTs by
the supplier in the form of repeated air oxidation at 400
$^{\circ}$C and acid washing results in purified and opened SWCNTs.
The tube
diameter distribution was determined from Raman spectroscopy \cite%
{KuzmanyEPJB} and we obtained $d=$ 1.40 nm and $\sigma $ = 0.10 nm
for the mean diameter and the variance of the Gaussian distribution,
respectively. Vapor filling with fullerenes was performed by
subjecting the SWCNT flakes to fullerene vapor in a sealed quartz
ampoule at 650 $^{\circ}$C following Ref. \cite{KatauraSM2001}.
Non-encapsulated fullerenes were removed by a 650 $^{\circ}$C
dynamic vacuum treatment. Typically 150 $\mu $g fullerenes in 100
$\mu $l solvent was sonicated for 1 h with 1 mg SWCNT in an
Eppendorf tube for the solvent peapod preparation. The weight uptake
of the SWCNT is $\sim $15 \% \cite{SimonPRL2005} that is shared
between the solvent and the fullerenes. The peapod was separated
from the solvent by centrifuging and the peapod material was greased
on a sapphire substrate. The vapor or solvent prepared peapods were
treated in dynamic vacuum at 1270 $^{\circ}$C for 2 hours for the
inner tube growth. The inner tube growth efficiency was found
independent of the speed of warming.

\textit{Raman spectroscopy.} Vibrational analysis was performed on a
Dilor xy triple Raman spectrometer in the 1.64-2.54 eV (676-488 nm)
energy range at ambient conditions.

\section{Results and discussion}

\begin{figure}[tbp]
\includegraphics[width=0.9\hsize]{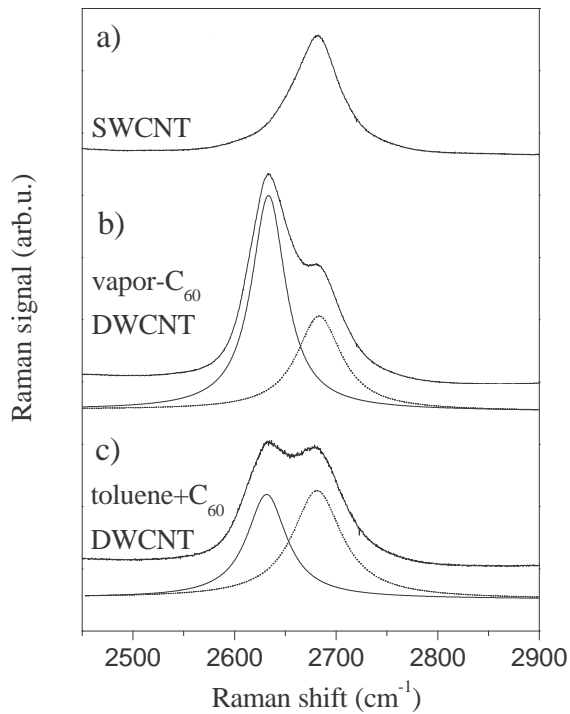}
\caption{The G' Raman mode of a) SWCNTs, b) vapor-C$_{60}$ peapod and c)
toluene+C$_{60}$ peapod based DWCNTs at $\protect\lambda $=515 nm laser
excitation (2.41 eV). Smooth solid and dashed curves show the deconvolution
to inner and outer tube modes, respectively.}
\label{toluene-vs-vapor-DWCNT}
\end{figure}

The growth of inner carbon nanotubes can be monitored by Raman
spectroscopy. We use extensively the response from the overtone of
the SWCNT D-line, known as the G' mode \cite{DresselhausTubesNew},
since this mode exhibits the largest absolute isotope shift. Fig.
\ref{toluene-vs-vapor-DWCNT} shows the G' mode for a SWCNT (a) and
two DWCNT samples that were obtained by high temperature annealing
from vapor prepared C$_{60}$ peapods (b) and toluene+C$_{60}$
peapods (c) by a 1270 $^{\circ}$C heat treatment. The vapor method
involves placing the SWCNT in a fullerene vapor in a sealed ampoule,
while the solvent method involves sonicating the solvent+C$_{60}$
solution together with the nanotubes which results in the
encapsulation of both the fullerene and the solvent
\cite{SimonCPL2004}. All results reported here are identical when
C$_{70}$ fullerenes were used. The G' mode provides a convenient
measure of the inner tube growth as the inner and outer tube
response are separated in the DWCNT samples: the mode on the low
Raman shifted side comes from the inner tubes
\cite{PfeifferPRB2005}. The ratio of the inner to outer tube mode
intensities strongly depends on the exciting laser energy with a
maximum around 2.41 eV \cite{PfeifferPRB2005}, that is used in the
current study.

\begin{figure}[tbp]
\includegraphics[width=0.9\hsize]{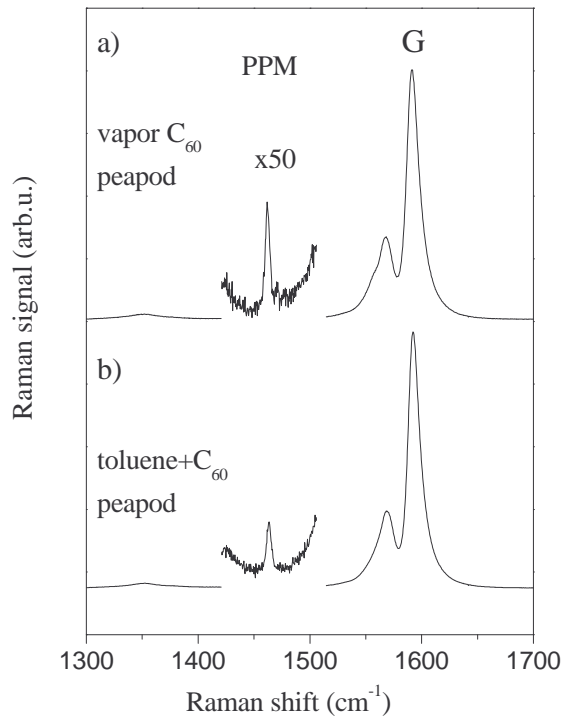}
\caption{The C$_{60}$ PPM mode and the nanotube G mode in a)
vapor-C$_{60}$ peapod and b) toluene+C$_{60}$ peapod sample at
$\protect\lambda $=488 nm laser excitation (2.54 eV). Note the 50
times magnified scale for the PPM range.}
\label{toluene-vs-vapor-peapod}
\end{figure}

The relative intensity of the inner tube mode in the vapor-C$_{60}$ and
solvent+C$_{60}$ peapod based DWCNTs measures the inner tube content in the
two samples. This can be quantified by deconvolution of the DWCNT signal
into inner and outer tube components as shown in Fig. \ref%
{toluene-vs-vapor-DWCNT}. The smaller inner tube signal in the solvent+C$%
_{60}$ sample is the result of a partial evaporation of the solvent before
the inner tubes are formed. It is known that encapsulation of the fullerene
is energetically preferred and thus it cannot escape from the tubes \cite%
{TomanekPRL}. However, a small molecule such as benzene and toluene can
leave the tube at higher temperatures. We found that the yield of inner
tubes grown from the benzene- (spectrum not shown) and toluene+C$_{60}$
peapods is 44(1)\ \% and 48(1) \% of that from the vapor-C$_{60}$ sample,
respectively. To separate the contribution to the inner tubes from the
solvents and C$_{60}$, the encapsulated C$_{60}$ content can be determined
in the solvent+C$_{60}$ peapod samples. In Fig. \ref{toluene-vs-vapor-peapod}%
, we show the spectral range that contains the C$_{60}$ pentagonal
pinch mode (PPM) at 1466 cm$^{-1}$ \cite{PichlerPRL2001} and the
nanotube G modes around 1650 cm$^{-1}$ \cite{DresselhausTubesNew}
for the vapor-C$_{60}$ and solvent+C$_{60}$ peapod sample. It was
demonstrated \cite{KuzmanyAPA} that the encapsulated C$_{60}$
content can be determined from the relative intensity of the PPM and
G modes. Assuming 100 \% filling of the available volume with
C$_{60}$ for the vapor-C$_{60}$ peapod sample \cite{LiuPRB2002}, the
results of Fig. \ref{toluene-vs-vapor-peapod} demonstrates only
40(1) \% fullerene filling for the toluene+C$_{60}$ sample. The same
value was obtained for benzene as a solvent.

In summary, from the 44 \% inner tube content in the benzene+C$_{60}$ peapod
based sample, $91$ \% of the carbon originates from C$_{60}$. Thus benzene
contributes only to approximately 9(4) \% of the carbon. Similarly, toluene
contributes to approximately 17(4)\ \% fraction of the total carbon amount
on the inner tubes. If we assume that solvents fill all available volume
apart from that filled with C$_{60}$, the current result means that a
significant portion, over 80 \%, is evaporated from both solvents without
contributing to the inner tube growth.

\begin{figure}[tbp]
\includegraphics[width=0.9\hsize]{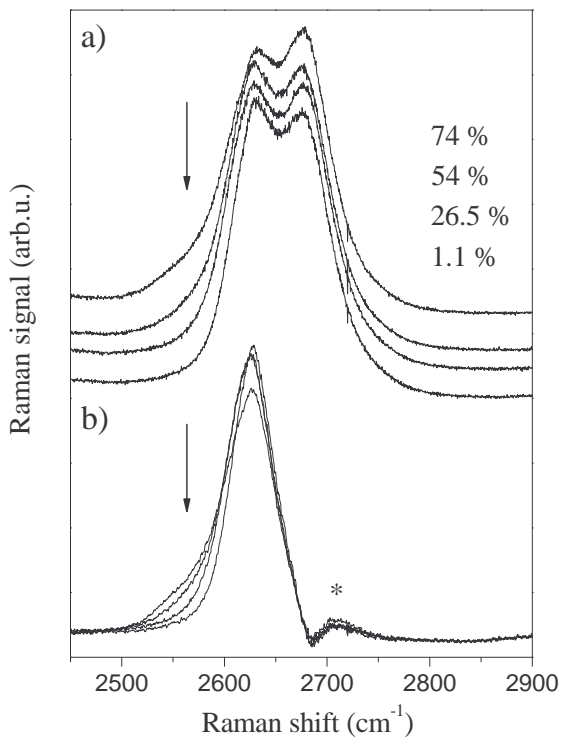}
\caption{a) The G' mode of toluene+C$_{60}$ peapod based DWCNTs with
varying $^{13}$C enrichment at $\protect\lambda $=515 nm laser
excitation (2.41 eV). From top to bottom: 74 \%, 54 \%, 26.5 \% and natural $%
^{13}$C content. b) The G' mode of the inner tubes after subtracting
the experimental SWCNT spectrum. Arrows indicate the spectral weight
shifted toward lower frequencies. A small residual peak is observed
around 2710 cm$^{-1}$ (denoted by an asterisk) due to the imperfect
subtraction.} \label{varying13C}
\end{figure}

Based on the current analysis, it can not be ruled out that e.g.
amorphous carbon is also inside the tubes or enters the tubes and
thus contributes to the inner tube growth. Therefore the crucial
statement on the growth of inner tubes from organic solvents
requires further reinforcement. To provide this, we studied the
inner tube growth from solvent+C$_{60}$ peapods where the solvent
was a mixture of $^{13}$C isotope labeled and natural solvents with
varying concentrations. Toluene was a mixture of ring $^{13}$C
labeled ($^{13}$C$_{6}$H$_{6}$-CH$_{3}$) and natural toluene
(C$_{7}$H$_{8}$). Benzene was a mixture of $^{13}$C enriched and
natural benzene. The labeled site was $>$ 99 \% $^{13}$C labeled for
both types of molecules. The $^{13}$C content, $x$, of the solvent
mixtures was calculated from the concentration of the two types of
solvents and by taking into account the
presence of the naturally enriched methyl-group for the toluene. In Fig. \ref%
{varying13C}a, we show the G' modes of DWCNTs with varying $^{13}$C
labeled content in toluene-C$_{60}$ based samples and in Fig.
\ref{varying13C}b, we show the same spectra after subtracting the
outer SWCNT component. A shoulder appears for larger values of $x$
on the low frequency side of the inner tube mode, whereas the outer
tube mode is unchanged. Similar behavior was observed for the
benzene+C$_{60}$ based peapod samples (spectra not shown) although
with a somewhat smaller spectral intensity of the shoulder. The
appearance of this low frequency shoulder is evidence for the
presence of a significant $^{13}$C content in the inner tubes. This
proves that the solvent indeed contributes to the inner tube
formation as it is the only sizeable source of $^{13}$C in the
current samples. The appearance of the low frequency shoulder rather
than the shift of the full mode indicates an inhomogeneous $^{13}$C
enrichment. A possible explanation is that smaller diameter
nanotubes might be higher $^{13}$C enriched as they retain the
solvent better than larger tubes.

\begin{figure}[tbp]
\includegraphics[width=0.9\hsize]{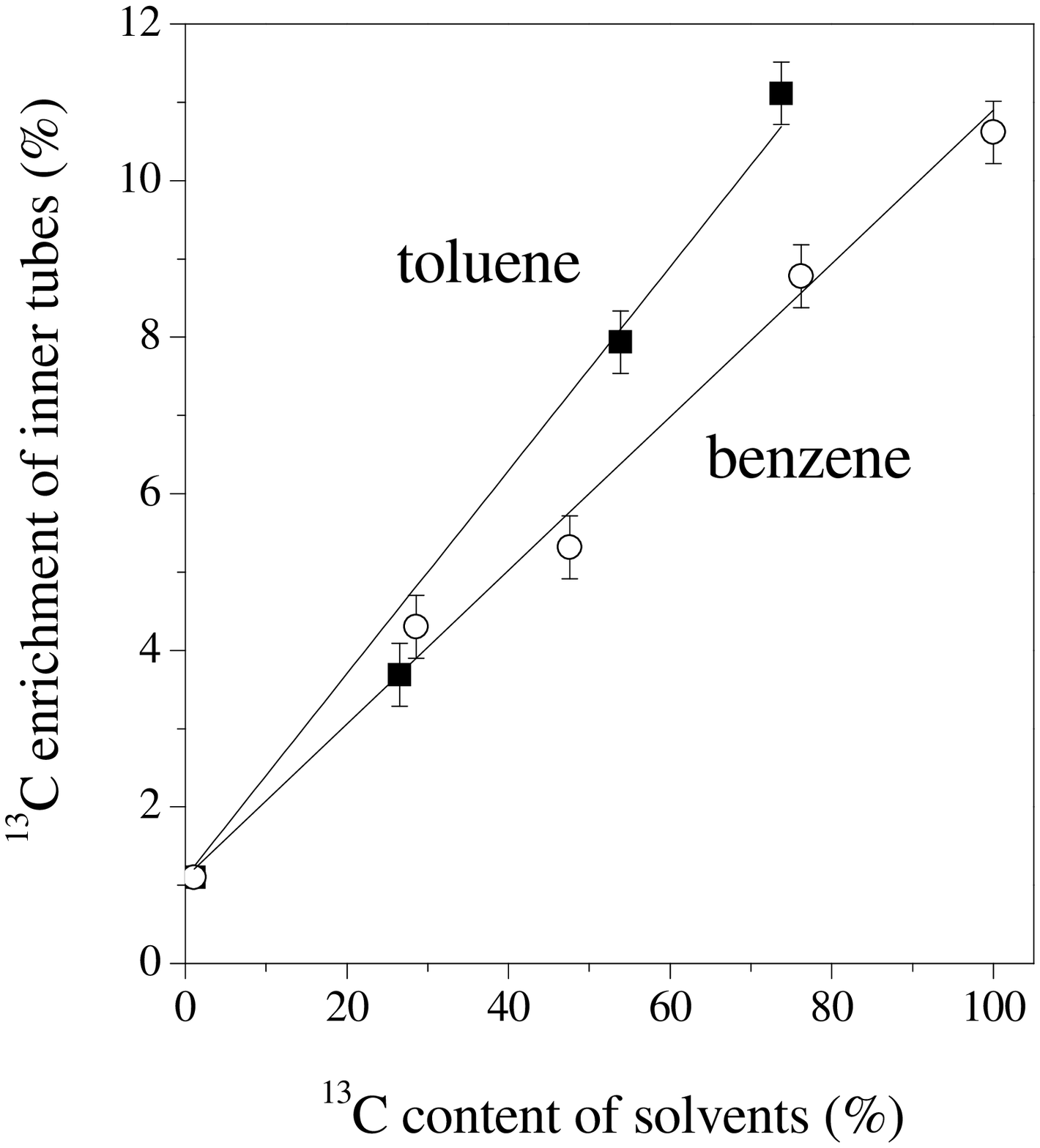}
\caption{$^{13}$C content of inner tubes based on the first moment analysis
as explained in the text as a function of $^{13}$C enrichement of benzene
and toluene. Lines are linear fits to the data as explained in the text. }
\label{13Cscaling}
\end{figure}

To quantify the $^{13}$C enrichment of the inner tubes, the
downshifted spectral weight of the inner tube G' mode was determined
from the subtracted spectra in Fig. \ref{varying13C}b. The
subtraction does not give a flat background above 2685 cm$^{-1}$,
however it is the same for all samples and has a small spectral
weight, thus it does not affect the current analysis. The
line-shapes strongly deviate from an ideal Lorentzian profile.
Therefore the line positions cannot be determined by fitting,
whereas the first moments are well defined quantities. The effective
$^{13}$C enrichment of the inner tubes, $c$, is calculated from
$\left( \nu _{0}-\nu \right) /\nu
_{0}=1-\sqrt{\frac{12+c_{0}}{12+c}}$, where $\nu _{0}$ and $\nu $
are the first moments of the inner tube G' mode in the natural
carbon and enriched materials, respectively, and $c_{0}=0.011$ is
the natural abundance of $^{13} $C in carbon. The validity of this
"text-book formula" was previously
verified by \textit{ab-initio} calculations for enriched inner tubes \cite%
{SimonPRL2005}. In Fig. \ref{13Cscaling}, we show the effective $^{13}$C
content in the inner tubes as a function of the $^{13}$C content in the
starting solvents. The scaling of the $^{13}$C content of the inner tubes
with that in the starting solvents proves that the source of the $^{13}$C is
indeed the solvents. The highest value of the relative shift for the toluene
based material, $\left( \nu _{0}-\nu \right) /\nu _{0}=0.0041(2)$,
corresponds to about 11 cm$^{-1}$ shift in the first moment of the inner
tube mode. The shift in the radial breathing mode range (around 300 cm$^{-1}$%
) \cite{DresselhausTubesNew} would be only 1 cm$^{-1}$. This
underlines why the
high energy G' mode is convenient for the observation of the moderate $^{13}$%
C enrichment of the inner tubes.

When fit with a linear curve with $c_{0}+A\ast x$, the slope, $A$ directly
measures the carbon fraction in the inner tubes that originates from the
solvents. The values, $A=0.098(5)$ and $A=0.130(6)$ agree with the solvent
related carbon fractions on the inner tubes of 9 and 17 \% as determined
from the intensity of the inner tube modes for benzene and toluene,
respectively.

The synthesis of inner tubes from organic solvent proves that any
form of carbon that is encapsulated inside SWCNTs contributes to the
growth of inner tubes. As mentioned above, inner tubes are not
formed in the absence of fullerenes but whether the fullerene is
C$_{60}$ or C$_{70}$ does not play a role. It suggests that
fullerenes act only as a stopper to prevent the solvent from
evaporating before the synthesis of the inner tube takes place. It
also clarifies that the geometry of fullerenes do not play a
distinguished role in the inner tube synthesis as it was originally
suggested \cite{SmalleyPRL2002}. It also proves that inner tube
growth can be achieved irrespective of the carbon source, which
opens a new prospective to explore the in-the-tube chemistry with
other organic materials.

\section{Acknowledgement}
FS acknowledges the Zolt\'{a}n Magyary programme for support. Work
supported by the Austrian Science Funds (FWF) project Nr. 17345, by
the EU projects MERG-CT-2005-022103 and BIN2-2001-00580, and by the
Hungarian State Grants (OTKA) No. TS049881, F61733 and NK60984.

$^{\ddag }$ Corresponding author, email: ferenc.simon@univie.ac.at,
Present address: Budapest University of Technology and Economics,
Institute of Physics and Solids in Magnetic Fields Research Group of
the Hungarian Academy of Sciences, H-1521, Budapest P.O.Box 91,
Hungary

\bibliographystyle{apsrev}

\bibliography{HabilNano}

\end{document}